\documentclass[a4paper, 11pt]{article}
\pdfoutput=1

\usepackage[british]{babel}
\usepackage[T1]{fontenc}
\usepackage[utf8]{inputenc}
\usepackage{jheppub}

\usepackage{booktabs}
\usepackage{capt-of}
\usepackage{xfrac}
\usepackage{mleftright}
\usepackage{csquotes}
\usepackage[noabbrev]{cleveref}

\newcommand{\beq}{\begin{equation}}
\newcommand{\eeq}{\end{equation}}
\newcommand{\bea}{\begin{eqnarray}}
\newcommand{\eea}{\end{eqnarray}}
\newcommand{\Lagr}{\mathcal{L}}

\hypersetup{
	pdftitle={MeV neutrino dark matter: Relic density, lepton flavour violation and electron recoil},
	pdfauthor={Juri Fiaschi, Michael Klasen, Miguel Vargas, Christian Weinheimer, Sybrand Zeinstra},
	pdfkeywords={Dark matter, neutrino masses, lepton flavour violation, direct detection}
}

\preprint{MS-TP-19-10}

\title{MeV neutrino dark matter: Relic density, lepton flavour violation and electron recoil}

\author[a]{Juri Fiaschi,}
\author[a]{Michael Klasen,}
\author[b]{Miguel Vargas,}
\author[b]{Christian Weinheimer,}
\author[a]{and Sybrand Zeinstra}

\affiliation[a]{Institut für Theoretische Physik, Westfälische Wilhelms-Universität
 Münster, Wilhelm-Klemm-Straße 9, 48149 Münster, Germany}
\affiliation[b]{Institut für Kernphysik, Westfälische Wilhelms-Universität
 Münster, Wilhelm-Klemm-Straße 9, 48149 Münster, Germany}

\emailAdd{fiaschi@uni-muenster.de}
\emailAdd{michael.klasen@uni-muenster.de}
\emailAdd{weinheimer@uni-muenster.de}
\emailAdd{m\_varg03@uni-muenster.de}
\emailAdd{swzeinstra@uni-muenster.de}

\abstract{%
Right-handed neutrinos with MeV to GeV mass are very promising candidates
for dark matter (DM). Not only can they solve the missing satellite puzzle,
the cusp-core problem of inner DM density profiles, and the too-big-to
fail problem, {\it i.e.} that the unobserved satellites are too big to not
have visible stars, but they can also account for the Standard Model (SM)
neutrino masses at one loop. We perform a comprehensive study of the
right-handed neutrino parameter space and impose the correct
observed relic density and SM neutrino mass differences and mixings.
We find that the DM masses are in agreement with bounds from big-bang
nucleosynthesis, but that these constraints induce sizeable DM couplings
to the charged
SM leptons. We then point out that previously overlooked limits from
current and future lepton flavour violation experiments such as MEG and SINDRUM
heavily constrain the allowed parameter space. Since the DM is leptophilic,
we also investigate electron recoil as a possible direct detection signal,
in particular in the XENON1T experiment. We find that despite the large
coupling and low backgrounds, the energy thresholds are still too high and
the predicted cross sections too low due to the heavy charged mediator,
whose mass is constrained by LEP limits.
}

\keywords{Dark matter, neutrino masses, direct detection, lepton flavour violation}

\begin{document}
\maketitle
\flushbottom


\section{Introduction}

While there is overwhelming evidence from many different length scales
for the existence of dark matter (DM) in the Universe, its nature is
still an unsolved problem in astroparticle physics \cite{Klasen:2015uma}.
Weakly Interacting Massive Particles (WIMPs) are very promising
candidates, since -- for masses in the GeV to TeV range and electroweak
coupling strength -- they lead in a straightforward way to the correct DM relic
density. The latter has now been measured by Planck with the very high
accuracy $\Omega h^2=0.1186\pm0.0020$, where $h$ denotes the present
Hubble expansion rate $H_0$ in units of 100 km s$^{-1}$ Mpc$^{-1}$
\cite{Aghanim:2018eyx}.

For many years, the lightest supersymmetric (SUSY) particle, the
lightest neutralino, stabilised by a discrete $R$-parity symmetry, has
been favoured, since there exist many other theoretical motivations for
SUSY \cite{Jungman:1995df}. The neutralino relic density
\cite{Herrmann:2007ku,Baro:2007em,Beneke:2014hja}, direct detection
\cite{Klasen:2016qyz} and LHC production cross sections \cite{Debove:2009ia}
can even be computed with a precision that rivals the experimental one.
However, neither the leading direct detection experiment in this mass
range, XENON1T \cite{Aprile:2018dbl}, nor the LHC \cite{epssusy} have
so far produced any evidence for SUSY DM.
In contrast, the discovery of a scalar boson of mass 125 GeV, consistent
with the properties of the Standard Model (SM) Higgs boson, by ATLAS
\cite{Aad:2012tfa} and CMS \cite{Chatrchyan:2012xdj} may imply the
existence of other scalars in Nature. When stabilised by a discrete
$Z_2$ symmetry similar to $R$-parity, the lightest neutral eigenstate
may well be the DM particle, as is the case {\it e.g.} in the inert
doublet model \cite{LopezHonorez:2006gr,Klasen:2013btp,Goudelis:2013uca}.

In addition to the evidence for DM, the observed solar and atmospheric
neutrino mass differences of $(7.59 \pm0.20)\cdot10^{-5}$ eV$^2$ and
$(2.43\pm0.13)\cdot10^{-3}$ eV$^2$ \cite{Tanabashi:2018oca} require
physics beyond the SM. Adding right-handed neutrinos to the SM has the
advantage that they do not only allow for neutrino mass generation
through different variants of the seesaw mechanism, but that they also
help to restore
parity at higher energy. If the seesaw mechanism is furthermore realised
at one loop, DM and neutrino masses become closely related as {\it e.g.}
in the scotogenic model \cite{Ma:2006km} or variants thereof
\cite{Restrepo:2013aga,Klasen:2013jpa}.
The detection prospects of DM depend not only on its spin and other
quantum numbers, but also strongly on its mass. While the
observed large scale structure of the Universe has long been believed to
favour the GeV to TeV cold DM regime \cite{Springel:2005nw}, warm DM, and
in particular keV sterile neutrinos, seem to account better for missing
satellite galaxies, the cusp-core problem of inner DM density
profiles and the too-big-to fail problem, {\it i.e.} the unobserved
satellites are too big to not have visible stars \cite{Adhikari:2016bei}.
Surprisingly, the MeV to GeV mass regime has so far remained largely
unexplored.

Notable exceptions are variants of the SLIM (Scalar as Light as MeV)
model, in which the SM is augmented at low energy by a scalar singlet
and right-handed neutrinos \cite{Boehm:2006mi} and in addition at
higher energy {\it e.g.} by a complex doublet \cite{Farzan:2009ji}.
Original motivation for this model came from the fact that it
could explain the 511 keV emission line from the centre of our galaxy,
observed by the INTEGRAL satellite \cite{Boehm:2003bt}.
If the scalar singlet is complex and almost as light as the lightest
sterile neutrino of MeV mass, the latter can be the DM and not only
lead approximately to the correct DM relic density and active neutrino
masses, but
also solve the missing satellite and too-big-to-fail problems
\cite{Arhrib:2015dez}. In addition to structure formation constraints,
this variant of the SLIM model has also passed collider constraints from LEP
and the LHC, that we will review below, and is thus not only very well
motivated, but also worthy of further investigation.

In this paper, we impose for the first time the precise Planck
measurement of the DM relic density as well as the neutrino mass differences
and mixing angles through the Casas-Ibarra parametrisation.
In addition, we study in detail the most competitive lepton flavour
violation constraints that had so far been overlooked in the
literature, but turn out to be highly restrictive on the allowed
parameter space due to the fact that the scalar doublet couples also
to the charged leptons \cite{Kubo:2006yx}.
Since the DM is light and leptophilic, {\it i.e.} has no tree-level
couplings to quarks, no visible signal from nuclear recoils is expected,
although the latter might well be induced at one loop for GeV to TeV DM
masses \cite{Kopp:2009et}. We therefore study instead the electron recoil
signal, which has recently come into focus for sub-GeV DM
\cite{Essig:2011nj,Lee:2015qva} and should not only be sensitive to
vector bosons
(dark photons), but also other mediators, in particular the scalar
doublet.

The goals of this paper are therefore threefold: First, after (re)defining
the SLIM model with MeV neutrino DM in Sec.\ \ref{sec:2} and reviewing in
Sec.\ \ref{sec:3} the theoretical and experimental constraints imposed
previously, we calculate in Sec.\ \ref{sec:4} the precise DM relic density
from neutrino mass difference and mixing constraints without any analytic
approximations. The result is not only the confirmation that neutrino
DM is indeed viable in the MeV to GeV mass range, but also the prediction
that its coupling to the scalar doublet can be large. Second, we compute
in Sec.\ \ref{sec:6} the expected lepton flavour
violation signals in the most sensitive channels $\mu\to e\gamma$,
$\mu\to3e$ and $\mu^-$Ti$\to e^-$Ti and compare them with the sensitivities
of current and future experiments. Third, we compute
in Sec.\ \ref{sec:5} the resulting electron recoil cross section and
compare it with new, realistic estimates of the XENON sensitivity and
other experiments under preparation.
In Sec.\ \ref{sec:7} we summarise our findings and discuss
their generalisation to other models with MeV-scale DM.


\section{The SLIM model with MeV neutrino DM}
\label{sec:2}

\subsection{Particle content}
\label{sec:2.1}

In the SLIM model, the SM scalar (Higgs) doublet $\Phi$ with mass
parameter $m_1$, self-coupling $\lambda_1$, squared vacuum expectation
value (VEV) $v^2=-2m_1^2/\lambda_1=(246\,{\rm GeV})^2$ and physical
Higgs boson mass $m_h^2=\lambda_1v^2=(125\,{\rm GeV})^2$ is augmented
at low energy by a real or complex scalar singlet. While both options
are in principle possible, we will discuss later the phenomenological
advantages of our choice of a complex scalar
\beq
 \rho = \frac{1}{\sqrt{2}} (\rho_R+i\rho_I)
\eeq
and (at least) two generations of singlet right-handed neutrinos $N_i$
in order to effectively explain the observed DM and two
non-zero neutrino masses at one loop \cite{Boehm:2006mi}. The
electroweak symmetry $SU(2)_L$ is restored at higher energy by the
addition of a complex scalar doublet \cite{Farzan:2009ji}
\beq
    \eta =\begin{pmatrix}
    \eta^+ \\
    \frac{1}{\sqrt{2}} (\eta_R+i\eta_I)
    \end{pmatrix}.
\eeq
All new particles are stabilised by a global $U(1)$ symmetry that is
softly broken to $Z_2$. At variance with the original scotogenic model,
which was directly based on the $Z_2$ symmetry and contained only the
complex scalar doublet and the right-handed neutrinos \cite{Ma:2006km},
the soft breaking of $U(1)$ and the mixing of the scalar singlet and the
neutral components of the scalar doublet, which both do not obtain a VEV,
allow for MeV neutrino DM and two similarly light neutral scalars and
thus for a solution to the missing satellite and too-big-to-fail problems
\cite{Arhrib:2015dez}. In the classification of Ref.\
\cite{Restrepo:2013aga}, the scotogenic and SLIM models correspond
to models T3-B and T1-1-A with hypercharge parameters $\alpha=-1$ and
0, respectively.

\subsection{Lagrangian}
\label{sec:2.2}

In the mass basis of the right-handed Majorana neutrinos $N_i$ ($i=1,2$) and
denoting the left-handed SM lepton flavour doublets as $L_j$ ($j=1,2,3$),
the Higgs potential and additional terms in the Lagrangian are given by
\cite{Farzan:2009ji,Arhrib:2015dez}
\bea
    \Lagr &=& - m_1^2 \Phi^\dagger \Phi - m_2^2 \eta^\dagger \eta - m_3^2 \rho^*\rho - \frac{1}{2} m_4^2 \left(\rho^2+ (\rho^*)^2 \right) - \mu (\eta^\dagger \Phi \rho + h.c.) - \frac{1}{2}m_{N_i} \overline{N^c_i}N_i \nonumber\\
    && - \frac{1}{2}\lambda_1 (\Phi^\dagger\Phi)^2 - \frac{1}{2}\lambda_2 (\eta^\dagger\eta)^2  - \frac{1}{2}\lambda_3 (\rho^*\rho)^2 - \lambda_4 (\eta^\dagger\eta)(\Phi^\dagger\Phi)- \lambda_5 (\eta^\dagger\Phi)(\Phi^\dagger\eta) \nonumber\\
    && - \lambda_6 (\rho^*\rho)(\Phi^\dagger\Phi)- \lambda_7 (\rho^*\rho)(\eta^\dagger\eta)  - \left(\lambda_8 \right)_{ij} (\overline{N^c_i} \eta^\dagger L_j + h.c.).
\label{eq:2.3}
\eea
This DM model is leptophilic, as can be seen from the last term involving
the couplings $\lambda_8$, and remains renormalisable despite the soft
breaking of the global $U(1)$ to $Z_2$, which implies that $m_4$ should
be small.

\subsection{Scalar masses and mixings after electroweak symmetry breaking}
\label{sec:2.3}

After electroweak symmetry breaking, the charged components of the
scalar doublet obtain the mass
\beq
 m_{\eta^\pm}^2 = m_2^2 + \frac{1}{2}\lambda_4 v^2.
 \label{eq:2.4}
\eeq
Due to the trilinear coupling $\mu$, the neutral components of the
complex singlet and doublet mix, and one obtains the mass matrix
\beq
    M_{R,I}^2 = \begin{pmatrix}
    m_2^2 + (\lambda_4 + \lambda_5) \frac{v^2}{2} & \mu \frac{v}{\sqrt{2}} \\
    \mu \frac{v}{\sqrt{2}} & m_3^2 + \lambda_6 \frac{v^2}{2} \pm m_4^2
    \end{pmatrix} =:
 \begin{pmatrix}A & \!\!\!\!B \\ B & \ C_{R,I} \end{pmatrix}
 \label{eq:2.5}
\eeq
with a positive (negative) sign of the $m_4^2$ term for the real (imaginary)
components. This leads to a very small mass splitting between the real and
imaginary scalar mass eigenstates
\beq
 \begin{pmatrix}\zeta_{1R,I} \\ \zeta_{2R,I}\end{pmatrix} =
 \begin{pmatrix} \cos{\theta_{R,I}}&-\sin{\theta_{R,I}}\\
 \sin{\theta_{R,I}}&\quad \cos{\theta_{R,I}}\\ \end{pmatrix}
 \begin{pmatrix}\eta_{R,I}\\ \rho_{R,I} \end{pmatrix},
\eeq
which are obtained by rotation about the angles $\theta_{R,I}$.
The eigenvalues of the mass matrix are given by
\beq
 m_{R,I}^2 = \frac{1}{2}\left(A+C_{R,I} \pm \sqrt{(A-C_{R,I})^2 +4B^2}\right).
 \label{eq:scalarmasses}
\eeq
Setting $B^2$ close to $AC_{R,I}$ with only a small difference
\beq
 AC_{R,I}-B^2=:\epsilon(A+C_{R,I})
 \label{eq:2.8}
\eeq
then allows to obtain two MeV neutral scalar mass eigenvalues
($m_{\zeta_{2R,I}}$), while the two others ($m_{\zeta_{1R,I}}$)
will be of the same size as or larger than $m_{\eta^\pm}$.


\section{Collider, cosmological and neutrino constraints}
\label{sec:3}

Although the SLIM model spans a relatively large parameter space, the
latter turns out to be restricted significantly both theoretically and
experimentally. Starting with the Higgs potential, both of its parameters
($m_1$ and $\lambda_1$) are now fixed by the known values of $v$ and
$m_h$ (see Sec.\ \ref{sec:2.1}). Theoretically, unitarity requires
certain combinations
of the couplings $\lambda_k$ ($k=2,...,8$) to lie below $8\pi$, while
vacuum stability and the need to avoid tachyonic particles require others
to be positive or larger than certain squared mass differences. These
theoretical constraints turn out to be almost automatically satisfied,
once collider and cosmological constraints are imposed. Several mass
parameters (in particular $m_{2,3}$) then also directly depend on
some of the couplings ($\lambda_{4,5,6}$).
An overview of all model parameters, as defined by the Lagrangian in
Eq.\ (\ref{eq:2.3}), together with their values or scan
ranges and phenomenological impact is given in Tab.\ \ref{tab:1}.
They are described in detail in the following.
\begin{table}
  \caption{\label{tab:1}Overview of model parameters with their values
  or scan ranges and phenomenological impact, respectively.}\vspace*{2mm}
  \centering
  \begin{tabular}{|c|c|c|l|}
    \hline
    Parameter   & Value & Range & Phenomenological impact \\
    \hline
    \hline
    $m_1^2$     &$-$(89 GeV)$^2$ & $-$     & Fixed by $v$ and $m_h$ \\
    $m_2$       & 83 GeV & $-$      & Fixed by $v$, $\lambda_4$ and $\lambda_5$ \\
    $m_3$       & 264 GeV     & $-$      & Fixed by $v$ and $\lambda_6$ \\
    $m_4$       & $-$   & 10 keV ... 10 MeV & Induces soft $U(1)$ breaking \\
    $\mu$       & $-$   &  251 GeV ... 252 GeV     & Fixed by $v$ and $\epsilon$ \\
    $\epsilon$  & $-$   & ($10^{-5}$ ... $6.1\cdot 10^1$) GeV$^2$ & Induces MeV masses \\
    $m_{N_1}$    & $-$   & (0.1 ... 0.98) $\cdot\,m_{\zeta_2}$ & ${\cal O}$(MeV) DM candidate \\
    $m_{N_2}$    & $-$   & 10 GeV ... 200 GeV & ${\cal O}$(GeV) sterile neutrino \\
    \hline
    $\lambda_1$ & 0.26  &  $-$     & Fixed by $v$ and $m_h$ \\
    $\lambda_2$ & 0.12  &  $-$  & Induces only scalar conversions \\
    $\lambda_3$ & 0.13  &  $-$  & Induces only scalar conversions \\
    $\lambda_4$ & 0.097 &  $-$  & Fixed by $v$, $m_{\eta^\pm}=99$ GeV, $R_{\gamma\gamma}$\\
    $\lambda_5$ & 0.13  &  $-$  & Fixed by $v$, $m_{\eta^\pm}=99$ GeV, $R_{\gamma\gamma}$\\
    $\lambda_6$ & 2.3   &  $-$  & Induces MeV masses \\
    $\lambda_7$ & 0.17  &  $-$  & Induces only scalar conversions \\
    $\lambda_8$ & $-$   & $10^{-6}$ ... $10^2$ & Fixed by Casas-Ibarra parametr.\\
    \hline
  \end{tabular}
\end{table}
%

\subsection{Collider constraints}
\label{sec:3.1}

The SM scalar Higgs doublet $\Phi$ couples to the new complex scalar
doublet $\eta$ and singlet $\rho$ through the matrix
\begin{equation}
 \begin{pmatrix} \eta\\ \rho \end{pmatrix}^{\dagger}
    \begin{pmatrix}
    (\lambda_4 + \lambda_5) v & \frac{\mu}{\sqrt{2}} \\
    \frac{\mu}{\sqrt{2}} & \lambda_6 v
    \end{pmatrix}
 \Phi
 \begin{pmatrix} \eta\\ \rho \end{pmatrix}.
\end{equation}
Inspection of Eq.\ (\ref{eq:2.5}) shows that, for negligibly small
$m_4$, the identifications   \cite{Arhrib:2015dez}
\bea
 m_2^2 = (\lambda_4 + \lambda_5) {v^2\over2} &\quad{\rm and}\quad&
 m_3^2 = \lambda_6 {v^2\over2}
 \label{eq:massids}
\eea
make the Higgs coupling matrix proportional to the scalar mass matrix.
While the invisible decays $h\to\zeta_{1R,I} \zeta_{1R,I}$ can be
forbidden by choosing $m_{\zeta_{1R,I}}>m_h/2$, the decays $h\to
\zeta_{2R,I}\zeta_{2R,I}\to N_1N_1\nu\nu$ will generally be kinematically
open due to the small values of $m_{\zeta_{2R,I}}$, but then also
have negligible branching ratios. As a consequence, $m_{2,3}$
are fixed by $\lambda_{4,5,6}$. Experimentally, the current ATLAS and
CMS limits on the invisible Higgs branching ratio lie at 26\% and
19\%, respectively \cite{Aaboud:2019rtt,Sirunyan:2018owy}.

A linear combination of the parameters $\lambda_{4,5}$ can in turn be
restricted by observing that the charged scalar mass $m_{\eta^\pm}$ in
Eq.\ (\ref{eq:2.4}) must lie above the rather model-independent LEP limit
of 98.5 GeV for charged scalars \cite{Abbiendi:2003yd}. The charged scalar
also contributes to the Higgs branching ratio into two photons, for which
CMS now measures a ratio to the SM value of $R_{\gamma\gamma}=
1.20^{+0.18}_{-0.14}$ \cite{Sirunyan:2018koj}. Choosing $\lambda_{4,5}
\simeq0.1$ $(\lambda_4 = 0.097, \lambda_5 = 0.13)$ satisfies both
constraints, as we have verified and can also be seen from Fig.\ 3 in
Ref.\ \cite{Arhrib:2015dez}. Larger values of $\lambda_4$ would lead to
larger $m_{\eta^\pm}$ and thus smaller electron recoil cross sections
and lepton flavour violation, while $\lambda_5$ has similar, but somewhat
less phenomenological influence.

\subsection{Cosmological constraints}
\label{sec:3.2}

Although studies of the SLIM model initially focused on the possibility
of {\em scalar} MeV DM \cite{Boehm:2006mi,Farzan:2009ji}, also to explain the
511 keV line from the galactic centre \cite{Boehm:2003bt}, it was later
noticed that the problems of missing satellite galaxies, the cusp
core of inner DM density profiles and the fact that the unobserved
satellites are too big to not have visible stars \cite{Adhikari:2016bei}
can rather be solved with MeV {\em neutrino} DM, provided that it scatters
off the active neutrinos in the early Universe by exchanging only slightly
heavier scalars \cite{Arhrib:2015dez}.

In the scalar sector, these cosmological constraints translate into
singlet and doublet mass parameters that should be close to each other,
so that the cancellation described in Sec.\ \ref{sec:2.3} can occur.
While $\lambda_{2,3,7}$ only control the scalar potentials and the
conversion of scalars into each other with little influence on the
phenomenology (we take {\it e.g.} $\lambda_2=0.12$, $\lambda_3=0.13$,
and $\lambda_7=0.17$), the singlet mass parameter $m_3$ should not be
much larger than $v$ ({\it i.e.} $\lambda_6\leq 3$), as was also
observed in Fig.\ 2 of Ref.\ \cite{Arhrib:2015dez}. Instead of varying
$\lambda_6$ and fixing $\epsilon$ to $10^{-4}$ GeV$^2$, we rather fix
$\lambda_6=2.3$ and vary the mass splitting parameter in Eq.\ (\ref{eq:2.8})
\beq
 \epsilon = (10^{-5}\,...\,6.1\cdot10^{1})\ {\rm GeV}^2.
\eeq
This has the advantage of more directly controlling the off-diagonal mass
parameter $\mu$ and thus the MeV mass of the two light scalars $\zeta_{2R,I}$.

It has been shown in Refs.\ \cite{Boehm:2000gq,Boehm:2004th}
that weak-strength DM interactions can erase the DM primordial
fluctuations when the DM is relatively light (of ${\cal O}$(MeV)) and coupled
to neutrinos or photons. The elastic scattering cross section must
then, however, be temperature-independent, which requires, in
addition to MeV DM mass and weak-strength DM couplings, a small mass
splitting of the MeV neutrino DM and the lightest scalar. This is
ensured in all our scenarios through the imposed scan ranges, in
particular of $m_4$, $\epsilon$ and $\lambda_6$.

For the lighter Majorana neutrino $N_1$ to represent DM and solve the
cosmological problems, it must therefore be (only somewhat) lighter than the
lightest scalar. We therefore vary $m_{N_1}/m_{\zeta_{2R,I}}$ in the
range $0.1\,...\,0.98$. In contrast, when the second Majorana neutrino
$N_2$ is heavy (10 GeV ... 200 GeV, the exact value being constrained by the
active neutrino mass differences and mixings), it decays promptly through
$N_2\to\eta \nu\to N_1\nu\nu$ and contributes neither significantly to the
relic density nor to any other signal considered here. A third Majorana
neutrino could of course be added, but would not significantly change the
phenomenology.

\subsection{Neutrino constraints}
\label{sec:3.3}

\begin{figure}[t!]
 \begin{center}
 \includegraphics[width=0.5\textwidth]{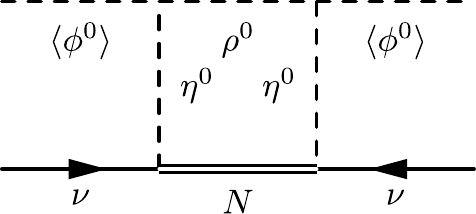}
 \end{center}
 \caption{One-loop diagram leading to the active neutrino mass matrix in
 the SLIM model.}
 \label{fig:1}
\end{figure}

In radiative seesaw models, the neutrinos obtain their masses only at the
loop level, and in particular in the SLIM model through the
one-loop diagram shown in Fig.\ \ref{fig:1}.

Taking the mixing in the neutral scalar sector as described in Sec.\
\ref{sec:2.3} into account, the neutrino mass matrix is given by
\cite{Arhrib:2015dez}
\bea
 (M_\nu)_{ij} &=&  \sum_k  \frac{(\lambda_8)_{ik}(\lambda_8)_{jk}}{16 \pi^2}m_{N_k} \left[\frac{\cos^2\theta_R m_{1R}^2}{m_{1R}^2-m_{N_k}^2}\ln{\frac{m_{1R}^2}{m_{N_k}^2}}  + \frac{\sin^2\theta_R m_{2R}^2}{m_{2R}^2-m_{N_k}^2}\ln{\frac{m_{2R}^2}{m_{N_k}^2}}      \right. \nonumber\\
    &&\left. - \frac{\cos^2\theta_I m_{1I}^2}{m_{1I}^2-m_{N_k}^2}\ln{\frac{m_{1I}^2}{m_{N_k}^2}}  - \frac{\sin^2\theta_I m_{2I}^2}{m_{2I}^2-m_{N_k}^2}\ln{\frac{m_{2I}^2}{m_{N_k}^2}}  \right].
\label{eq:3.4}
\eea
As in the scotogenic model, the neutrino masses arise from the small mass
differences between the real and imaginary scalars. In the SLIM model, this
mass splitting is realised by the non-zero value of $m_4$ (cf.\ Sec.\
\ref{sec:2.3}), required to be
small from the soft breaking of $U(1)$ to $Z_2$. We do not fix $m_4$ here
to the relatively large value of 3 MeV as in Ref.\ \cite{Arhrib:2015dez},
but allow it to vary in the range 10 keV ... 10 MeV, as a smaller mass
splitting of $\zeta_{2R,I}$ reduces the mass-dependent terms in the neutrino
mass matrix and thus indirectly allows for larger lepton couplings
$\lambda_8$.

We take the neutrino mass and mixing constraints into account assuming
for simplicity a massless lightest neutrino, which implies a normal
hierarchy and absolute values for the other two neutrino masses in the
mass matrix
\beq
 D_\nu = U_\nu^\dagger M_\nu U_\nu = \text{diag}(0, m_{\nu 2}, m_{\nu 3}),
 \label{eq:3.5}
\eeq
diagonalised by the PMNS matrix $U_\nu$.

To impose the experimental constraints on the neutrino mass and mixing, we
make use of the Casas-Ibarra parametrisation \cite{Casas:2001sr}. It takes
as an input the experimental neutrino data and the dark particle masses and
mixings and returns the coupling $\lambda_8$. Specifically, we rewrite Eq.
\eqref{eq:3.4} as
\beq
    M_\nu = \lambda_8^T M \lambda_8,
\eeq
where $M$ is a $3\times 3$ diagonal matrix whose elements are differences
of mass functions, summed over $k$. Together with Eq.\ (\ref{eq:3.5}) we
then obtain
\beq
 D_\nu = U_\nu^\dagger \lambda_8^T M \lambda_8 U_\nu
\eeq
or
\beq
 \text{diag}(0,1,1) = D_\nu^{-1/2} U_\nu^\dagger \lambda_8^T M \lambda_8 U_\nu D_\nu^{-1/2} \equiv R^\dagger R,
\eeq
where the rotation angle $\theta$ in
\beq
R = \begin{pmatrix}
0 & \cos \theta & \sin \theta \\
0 & -\sin \theta & \cos \theta
\end{pmatrix}
\eeq
is arbitrary, {\it i.e.} it can vary in the range 0 ... $2\pi$.
Identifying
\beq
 R = M^{1/2} \lambda_8 U_\nu D_\nu^{-1/2}
\eeq
then leads to the desired couplings
\beq
 \lambda_8 = M^{-1/2} R D_\nu^{1/2} U_\nu^\dagger,
\eeq
that are consistent with the neutrino data and depend not only on
$\theta$, but also indirectly on the scalar and right-handed
neutrino masses as well as the scalar couplings in the potential.
The coupling $\lambda_8$ itself determines many
properties of the model. Apart from being responsible for the
neutrino masses, it also directly influences the relic density,
the electron recoil cross section, as well as the amplitude of
lepton flavour violating processes such as $\mu \rightarrow e
\gamma$.


\section{Right-handed neutrino DM relic density, mass and lepton couplings}
\label{sec:4}

\begin{figure}[t!]
 \centering
 \includegraphics[width=\textwidth]{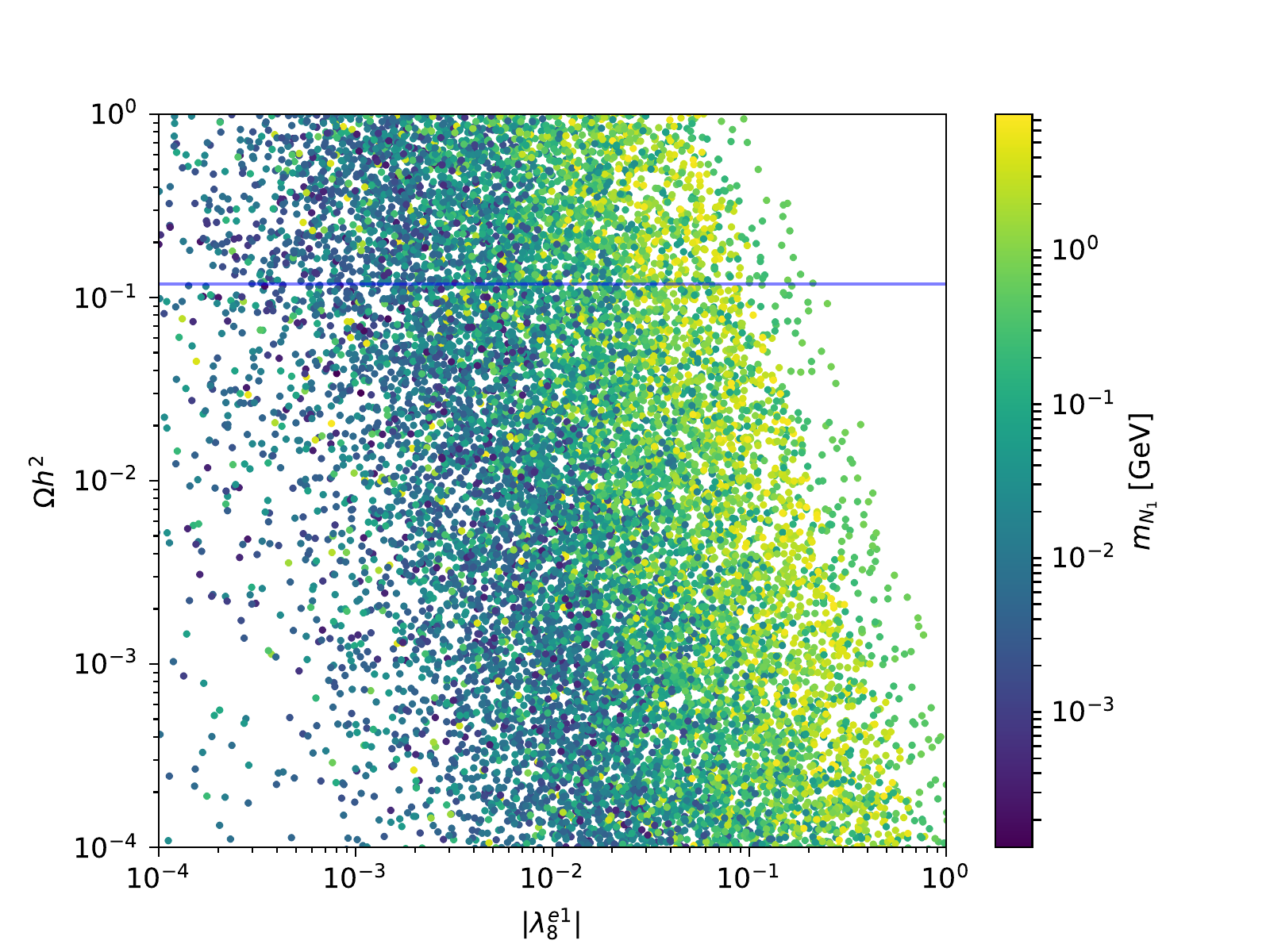}
 \caption{The relic density $\Omega h^2$ obtained in the SLIM model with
 right-handed neutrino DM versus the absolute value of its coupling to
 electrons and electron neutrinos
 $\lambda_8^{e1}$. The relic density measured by Planck \cite{Aghanim:2018eyx}
 is shown as a blue horizontal band, and the lightest Majorana neutrino mass
 $m_{N_1}$ is given on a colour scale.}
\label{fig:2}
\end{figure}

We now come to the first new result of our study, {\it i.e.} a numerical
calculation of the precise right-handed neutrino DM relic density without
any analytic approximations. To this end, we reproduce the Lagrangian of
the SLIM model \cite{Farzan:2009ji,Arhrib:2015dez} presented in Eq.\
(\ref{eq:2.3}) from the particle content defined in Sec.\ \ref{sec:2.1}
using our program \textsc{minimal-lagrangians} \cite{may}. The corresponding
Feynman rules are then derived with \textsc{SARAH 4.14.2} \cite{Staub:2013tta}
and passed on to \textsc{SPheno 4.0.3} \cite{Porod:2011nf} to calculate the
physical mass spectrum for each point in the parameter space that we have
defined in Sec.\ \ref{sec:3}.

At this point, we do not only impose the collider, cosmological and
neutrino constraints in the ways described there, but also use
\textsc{SPheno 4.0.3} to check (again) other collider and low-energy
constraints such as the $Z$ and Higgs boson branching ratios, the $\rho$
parameter, and the braching ratio of the flavour-changing neutral current
process $b\to s\gamma$. The right-handed neutrino DM relic density for each
surviving model is then calculated with \textsc{micrOMEGAs 4.3.5}
\cite{Barducci:2016pcb}.

Figure \ref{fig:2} shows the DM relic density $\Omega h^2$ of all viable
SLIM models with right-handed neutrino DM versus the absolute value of
its coupling to electrons and electron neutrinos $\lambda_8^{e1}$,
coloured according to the value of the mass $m_{N_1}$ of the DM candidate.
One can clearly see that the relic density drops quickly from $10^0$ to
$10^{-4}$ when the lepton coupling rises by about two orders of magnitude,
{\it e.g.} from a few times $10^{-4}$ to $10^{-2}$ or $10^{-2}$ to a few
times $10^{-1}$. However, the correct relic density as observed by Planck
(blue horizontal band) \cite{Aghanim:2018eyx} can be obtained for all couplings
in these ranges, provided that the DM mass rises simultaneously from the
MeV (dark blue) to the GeV scale (light green/yellow). We have therefore
confirmed that while MeV dark matter is indeed consistent with collider,
structure formation and neutrino constraints, the correct relic density
imposes further restrictions and in fact predicts sizeable
couplings of almost up to ${\cal O}(1)$ to the SM leptons. In the following,
we therefore impose also the Planck value for the relic density.

\begin{figure}[t!]
 \centering
 \includegraphics[width=\textwidth]{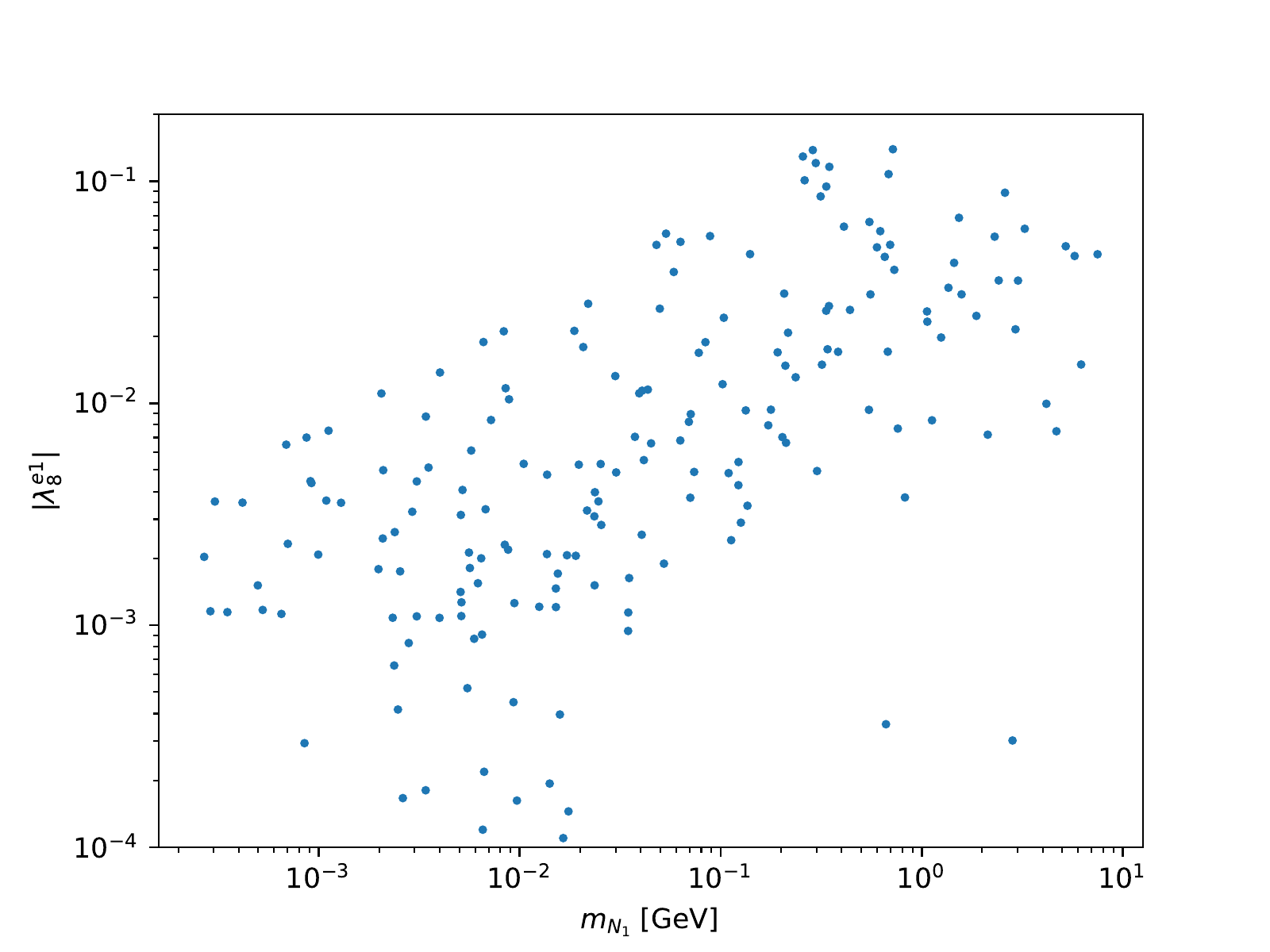}
 \caption{DM-electron coupling $\lambda_8^{e1}$ versus DM mass $m_{N_1}$
   for models which
   satisfy not only collider, cosmological and neutrino mass constraints, but
   also lead to the correct relic density.}
\label{fig:9}
\end{figure}

In Fig.\ \ref{fig:9} we show all models, which satisfy not only collider,
cosmological and neutrino mass constraints, but also lead to the correct
relic density. It therefore represents a projection of Fig.\ \ref{fig:2}
on the blue line and is displayed in the plane of the absolute value of
the DM-electron coupling $|\lambda_8^{e1}|$ versus DM mass $m_{N_1}$. As
one can observe, heavier DM implies enhanced DM annihilation processes
through larger couplings to the SM leptons in order to reduce the DM
abundance and still obtain the correct relic density.
%


\section{Lepton flavour violation and experimental constraints}
\label{sec:6}

\begin{figure}[ht]
 \begin{center}
 \includegraphics[width=0.9\textwidth]{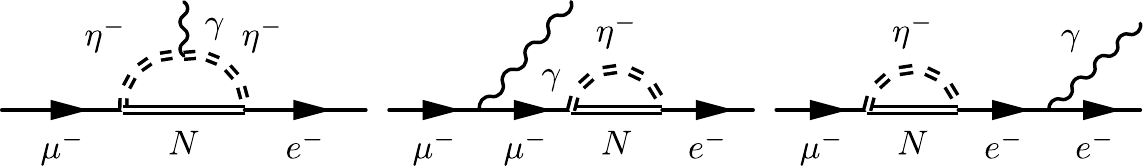}
 \end{center}
 \caption{Feynman diagrams contributing to the lepton flavour violating
 process $\mu \rightarrow e \gamma$ at one loop through the exchange of
 $Z_2$-odd right-handed neutrinos ($N$) and charged scalars ($\eta^-$).}
 \label{fig:5}
\end{figure}

Since the right-handed neutrinos $N_i$ in our model couple through the
complex scalar doublet $\eta$ not only to SM neutrinos, but also to the
charged leptons,
processes that violate lepton flavour naturally occur at the one-loop
level. The most sensitive process with the strictest experimental bounds is
usually the flavour-changing neutral current process $\mu \rightarrow e
\gamma $. Here, it occurs through the Feynman diagrams shown in Fig.\
\ref{fig:5}
 and depends, apart from the exchanged particle masses,
not only on the couplings $\lambda_8^{ei}$, but also on $\lambda_8^{\mu i}$,
which are both constrained by the neutrino masses and mixings through the
Casas-Ibarra parametrisation.

\begin{figure}[t]
 \centering
 \includegraphics[width=0.9\textwidth]{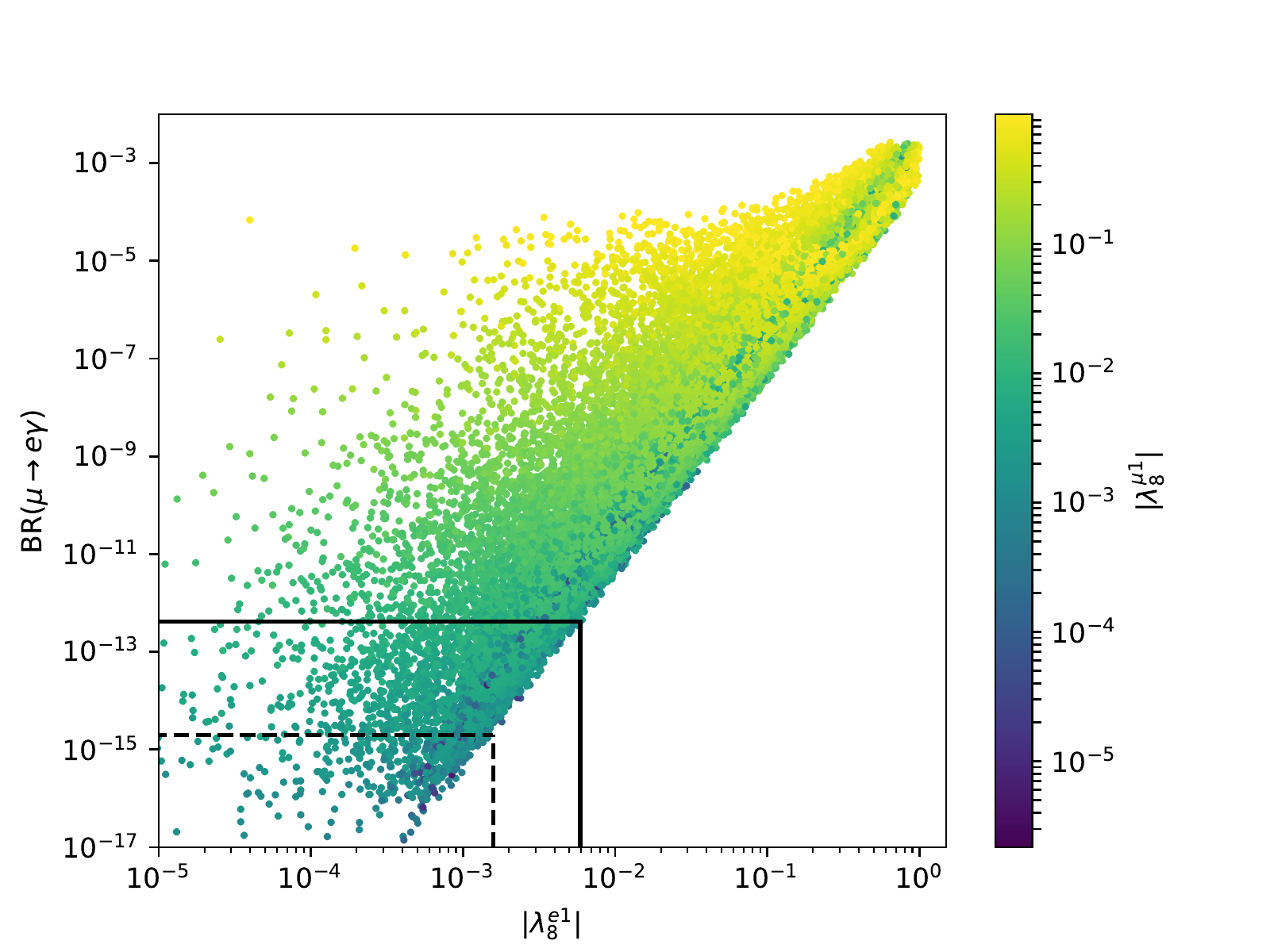}
 \caption{Branching ratio of the lepton flavour violating process
 $\mu \rightarrow e \gamma$ versus the DM-electron coupling $\lambda_8^{e1}$.
 The DM-muon coupling $\lambda^{\mu1}_8$ is shown on a colour scale.
 Also shown are the current (solid black line) \cite{TheMEG:2016wtm}
 and expected future (dashed black line) \cite{Renga:2018fpd} limits on the
 branching ratio by the MEG experiment.}
 \label{fig:6}
\end{figure}

We compute the branching ratio of the process $\mu\to e\gamma$
 with \textsc{SPheno 4.0.3} \cite{Porod:2011nf}
and plot it in Fig.\ \ref{fig:6} as a function of the dominating couplings
$\lambda_8^{e1}$ on the $x$-axis and $\lambda_8^{\mu 1}$ on a colour scale.
Interestingly, the non-zero neutrino masses impose a lower bound on this
branching ratio of about $10^{-17}$. Experimentally, the MEG experiment
currently excludes branching ratios above $4.2 \cdot 10^{-13}$
\cite{TheMEG:2016wtm}. This already excludes a substantial part of the
parameter space and in particular couplings $\lambda_8^{e1}$ above
$6\cdot 10^{-3}$ and $\lambda_8^{\mu1}$ above about $10^{-2}$. In the
SLIM model with scalar DM, similar limits of a few times $10^{-3}$
and a few times $10^{-2}$, respectively, have been expected in an approximate
calculation \cite{Farzan:2009ji}.
After the planned upgrade of the MEG experiment, it is expected
to reach a sensitivity of $2 \cdot 10^{-15}$ \cite{Renga:2018fpd}, which
would lower the limits on the couplings by about a factor of four, leaving
open only a small part of the parameter space.

\begin{figure}[t]
 \centering
 \includegraphics[width=0.9\textwidth]{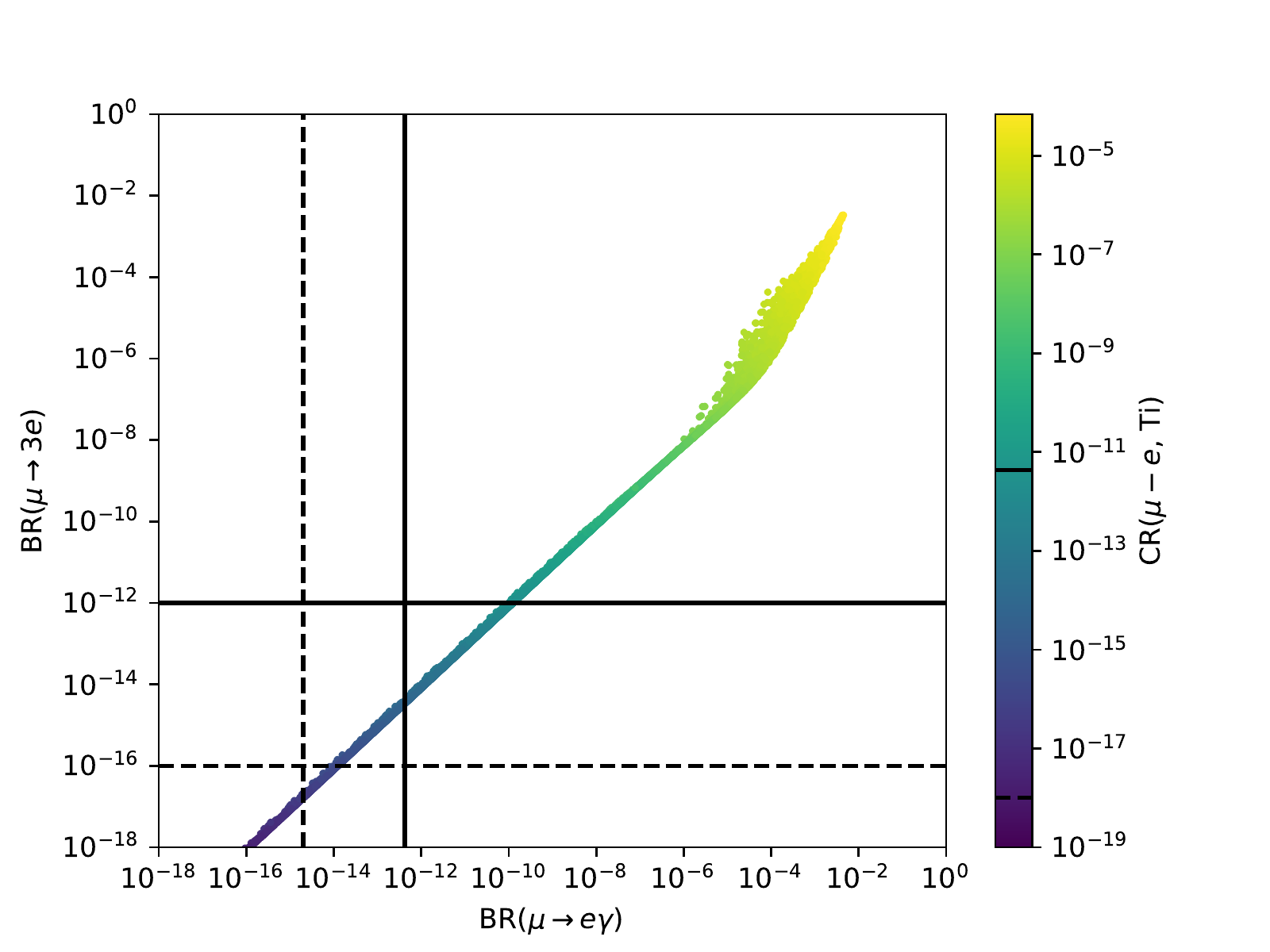}
 \caption{Branching ratio of the lepton flavour violating process
   $\mu \rightarrow 3e$ versus the one of the (usually most constraining)
   process $\mu \rightarrow e \gamma$. The rate of muon conversion in the
   field of titanium atoms $\mu^-$Ti $\to e^-$Ti is shown on a colour scale.
   Also shown are the current (solid black lines) \cite{TheMEG:2016wtm,
   Bellgardt:1987du,Dohmen:1993mp} and expected future (dashed black lines)
   \cite{Renga:2018fpd,Blondel:2013ia,Sato:2008zzm} experimental limits on
   the branching ratios and conversion rate.}
 \label{fig:8}
\end{figure}

The current limit on the branching ratio $\mu\to3e$ obtained by the
SINDRUM experiment lies at 10$^{-12}$ \cite{Bellgardt:1987du}. It is thus
only slightly weaker than the MEG limit on $\mu\to e\gamma$. In addition,
SINDRUM II obtained a limit of $4.3\cdot10^{-12}$ on the muon-to-electron
conversion rate in Titanium \cite{Dohmen:1993mp}. Significant progress is
expected for both processes with possible future limits of 10$^{-16}$
\cite{Blondel:2013ia} and even $10^{-18}$ \cite{Sato:2008zzm},
respectively. In Fig.\ \ref{fig:8} we therefore show predictions for
these processes in our viable models, satisfying all collider, cosmological
and neutrino constraints, together with those for $\mu\to e\gamma$
and compare them to the current (solid lines) and future (dashed lines)
experimental limits. We observe a strong correlation of all three processes,
except for large, already excluded branching ratios and conversion rates.
While the current limit on $\mu\to e\gamma$ sets the strongest bound, the
other two experiments will indeed reach a similar sensitivity in the future.
Note that the scans in Figs.\ \ref{fig:6} and \ref{fig:8} do not include the
constraint on the correct relic density, i.e.\ also points which under- (or
over-) saturate the relic density are included.


\section{Electron recoil cross section and experimental sensitivity}
\label{sec:5}

As described in the introduction, the DM in the SLIM model is leptophilic,
so that one does not expect any nuclear recoil signal from experiments
like XENON1T, which has set the current best limit on the spin-independent
direct detection cross section for DM masses between 6 GeV and 1 TeV
\cite{Aprile:2018dbl}, XENON100, which had extended the sensitivity down
to 3.7 GeV using the ionisation (S2) signal only \cite{Aprile:2016wwo}, or
CRESST-III, whose cryogenic CaWO$_4$ crystals are even sensitive down to
DM masses of 160 MeV \cite{Abdelhameed:2019hmk}. 

Given that the DM is of MeV to GeV mass and its couplings to the SM
leptons are large, one would, however, expect an observable electron
recoil signal, which has recently come into focus for sub-GeV DM and
should not only be sensitive to vector bosons (dark photons), but also
other mediators \cite{Essig:2011nj,Lee:2015qva}.
If we generalise the notation for the DM particle to the conventional
$\chi$, the ionisation rate, differential in the electron recoil (er)
energy $E_\text{er}$,
\beq
 \frac{dR_{\text{ion}}}{d\ln E_{\text{er}}} =
 N_{\text{T}} \frac{\rho_{\chi}}{m_{\chi}}
 \sum_{n\ell} \frac{d\langle \sigma_{\text{ion}}^{n\ell}v \rangle }
 {d\ln E_{\text{er}}}
 \label{eq:rate}
\eeq
is proportional to the number of target nuclei per unit mass $N_T$,
the local DM density $\rho_\chi= 0.4$ GeV/cm$^3$ \cite{Read:2014qva},
the differential thermally averaged cross section
\beq
 \frac{d\langle \sigma_{\text{ion}}^{n\ell}v  \rangle }{d\ln E_{\text{er}}} =
 \frac{\bar{\sigma}_{\chi \text{e}}}{8 \mu_{\chi \text{e}}^2}
 \int \ | f_{\text{ion}}^{n\ell}(k',q) |^2 F(k',Z_{\rm eff})
 | F_{\text{DM}}(q)|^2 \eta(v_{\text{min}},t) \ q \ dq,
 \label{eq:5.2}
\eeq
summed over all possible electron states $(n,\ell)$, and inversely
proportional to the DM mass $m_\chi$. In Eq.\ (\ref{eq:5.2}),
$\mu_{\chi \text{e}}$ is the DM-electron reduced mass, and $k'=\sqrt{2m_e
E_\text{er}}$ and $q$ are the outgoing electron momentum and momentum
transfer, respectively. Assuming plane waves for the scattered electrons
and a spherically symmetric atom with full shells, the form factor for
ionisation reduces to 
\beq
 \Big| f^{n \ell}_{\text{ion}}(k',q) \Big|^2 =  \frac{(2\ell+1)k'^2}{4\pi^3 q}  \int \big| \chi_{n\ell}(k) \big|^2 k dk,
 \label{eq:ff}
\eeq
where $\chi_{n \ell}(k)$ is the radial part of the momentum space
wave function of the bound electron \cite{Kopp:2009et}. As in the
case of nuclear beta decay, the wave function of the escaping electron
is distorted by the presence of the nearby atom, requiring that the
rate be corrected by the Fermi function
\bea
 F(k',Z_{\rm eff})={2\pi\nu\over1-e^{-2\pi\nu}} &\quad {\rm with} \quad &
 \nu=Z_{\rm eff}{\alpha m_e\over k'}.
\eea
$Z_{\rm eff}$ is the effective charge felt by the electron, expected
to be somewhat larger than unity. However, for outer shell electrons
$Z_{\rm eff}=1$ is a good approximation \cite{Essig:2011nj,Lee:2015qva}.
Treating the electrons in the target as single-particle states of an
isolated atom, we can use the tabulated numerical Roothaan-Hartree-Fock
(RHF) wave functions \cite{Bunge:1993jsz}. Astrophysics enters through
the mean inverse DM velocity \cite{Lee:2015qva}
\beq
 \eta(v_{\text{min}},t) = \int_{v_{\min}}^{\infty}
 \frac{d^3v}{v} f({\bf v},t),
 \label{eq:eta}
\eeq
which depends on the Earth-frame velocity distribution of DM $f({\bf v},t)$,
that acquires a time dependence as the Earth orbits the Sun.
In the Galactic frame, and asymptotically far away from the Sun’s
gravitational potential, we take the velocity distribution
$f_\infty({\bf v})$ to be that of the Standard Halo Model
\beq
f_\infty({\bf v})=1/N_{\rm esc}(\pi v_0^2)^{-3/2} e^{-{\bf v}^2/v_0^2},
 \quad {\rm if}\quad |{\bf v}|\leq v_{\rm esc},
\eeq
where $N_{\rm esc}$ is a normalisation
factor. We adopt a local circular velocity of $v_0$ = 220 km/s and an
escape velocity of
$v_{\text{esc}}$ = 544 km/s \cite{Smith:2006ym}. The minimal DM velocity
\begin{equation} \label{eq:vmin}
 v_{\min} = \frac{E_{\text{B}}^{n\ell} + E_{\text{er}}}{q} + \frac{q}{2m_{\chi}}
\end{equation}
depends on the required sum of binding energy $E_{\text{B}}^{n\ell}$ and recoil
energy $E_{\text{er}}$.

\subsection{Theoretical expectations}

\begin{figure}[t]
 \begin{center}
 \includegraphics[width=0.9\textwidth]{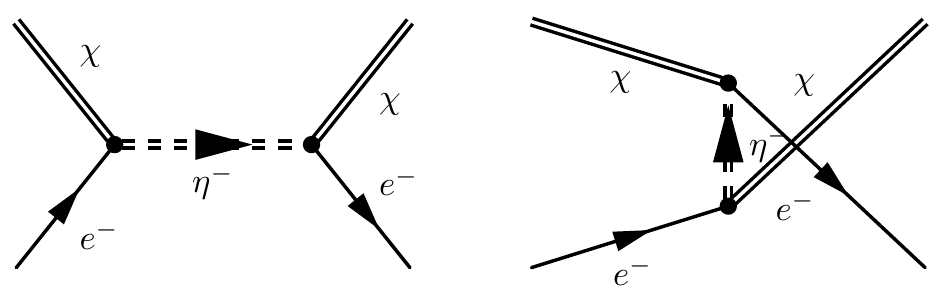}
 \end{center}
 \caption{Feynman diagrams contributing to the scattering of right-handed
 neutrino DM ($\chi=N_1$) off electrons ($e^-$) through $s$- and $u$-channel
 exchanges of the charged $Z_2$-odd scalars ($\eta^-$).}
 \label{fig:3}
\end{figure}

In Eq.\ (\ref{eq:5.2}), we are of course primarily interested in the
DM scattering cross section off electrons. It is usually factorised
into the non-relativistic reference cross section
\beq
 \bar{\sigma}_{\chi \text{e}}  = \frac{\mu_{\chi \text{e}}^2 }{16\pi m_{\chi}^2m_{\text{e}}^2}  \overline{ |\mathcal{M}_{\text{e}\chi}(q)|^2}\Big|_{q^2 = \alpha^2 m_{\text{e}}^2}
 \label{eq:sigma}
\eeq
at fixed momentum transfer $q=\alpha m_{\text{e}}$, where $\alpha$ is
the electromagnetic fine structure constant and $m_e$ is the electron mass,
and the form factor
\begin{equation} \label{eq:dmff}
|F_{\text{DM}}(q) |^2 = \overline{|\mathcal{M}_{\text{e}\chi}(q)|^2}/ \overline{|\mathcal{M}_{\text{e}\chi}(\alpha m_{\text{e}})|^2},
\end{equation}
that captures the $q$-dependence of the matrix element. For light
mediators, $F_{\text{DM}}(q)=\alpha^2m_e^2/q^2$, whereas for heavy
mediators $F_{\text{DM}}(q)=1$.
In our model, the DM particles $\chi$ are the right-handed neutrinos $N_1$.
They scatter off the electrons through the exchange of the $Z_2$-odd
scalars $\eta^-$, coupling with strength $\lambda_8^{e1}$, in the $s$-
and $u$-channel, as shown in the Feynman diagrams in Fig.\ \ref{fig:3}.
Since the charged scalar $\eta^-$ is heavy (cf.\ Sec.\ \ref{sec:3.2}),
the propagator can be integrated out with the result that the form
factor $F_{\text{DM}}(q)=1$ and the reference cross section is
\bea
 \bar{\sigma}_{\chi \text{e}} &=& \frac{\mu_{\chi e}^2 (\lambda_8^{e1})^4}{ \pi m_{\eta^\pm}^4}.
 \label{eq:5.9}
\eea

\begin{figure}[h!]
 \centering
 \includegraphics[width=\textwidth]{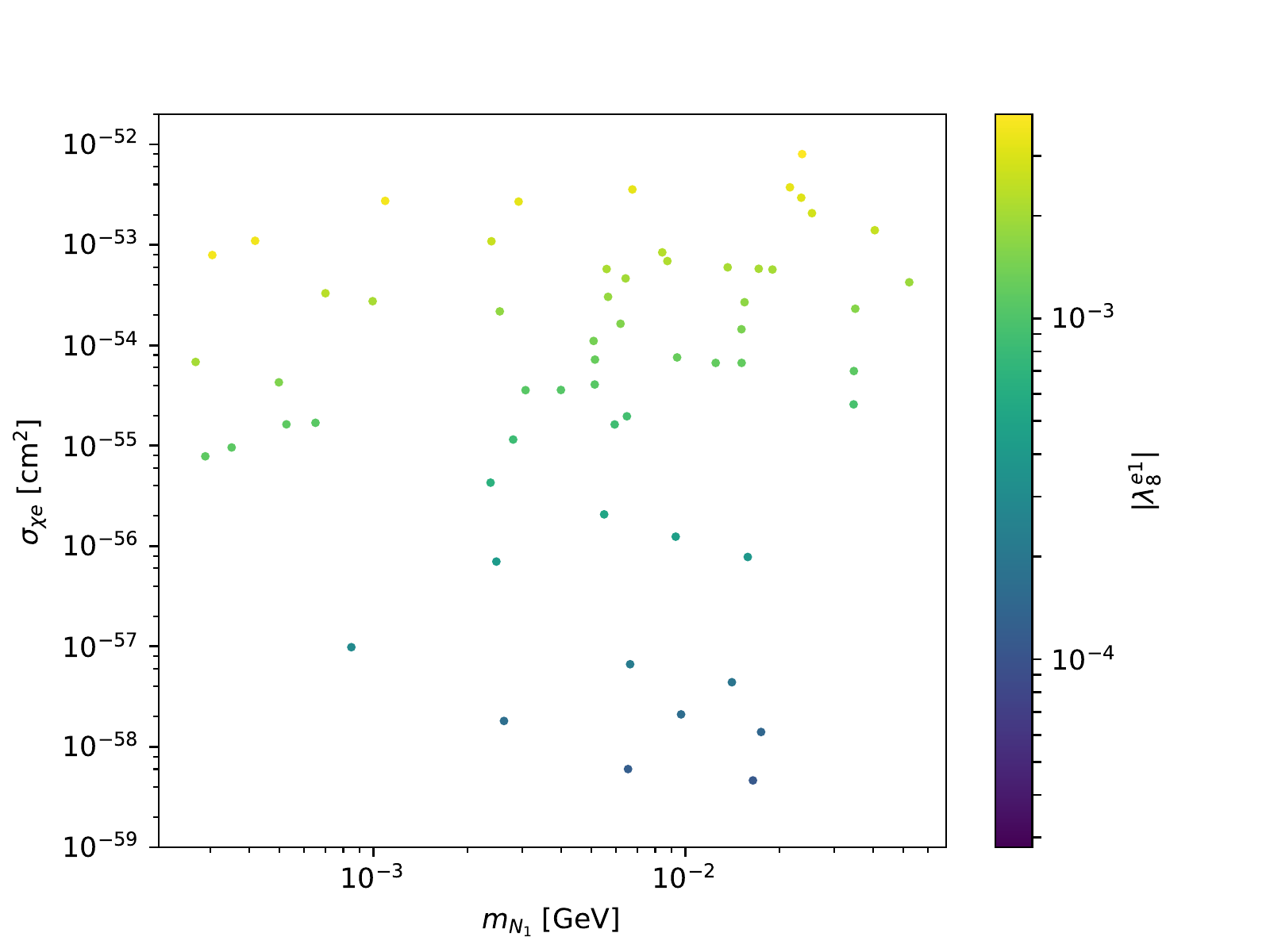}
 \caption{Scattering cross section of right-handed neutrino DM off
 electrons versus the DM mass. The DM-electron coupling $\lambda_8^{e1}$
 is shown on a colour scale.}
 \label{fig:4}
\end{figure}

For the models satisfying all the constraints described in Sec.\ \ref{sec:3}
and reproducing the observed DM relic density as
described in Sec.\ \ref{sec:4}, the numerical values of the scattering
cross sections of right-handed neutrino DM off electrons are shown in
Fig.\ \ref{fig:4} as a function of the DM mass. In addition, the size of
the DM-electron coupling $\lambda_8^{e1}$ is shown on a colour scale.
The cross sections do not exceed a few times $10^{-46}$
cm$^2$ even for the largest couplings $\lambda_8^{e1}\geq 0.1$
and reach only $10^{-52}$ cm$^2$, when in addition the lepton flavour
violation constraints of Sec.\ \ref{sec:6} are imposed.
The reason is the strong suppression by the large mass $m_{\eta^\pm}$ of
the charged scalar mediator, which is constrained by the LEP limit to be
heavier than 98.5 GeV (cf.\ Sec.\ \ref{sec:3.1}) and which enters with
the fourth power in Eq.\ (\ref{eq:5.9}).
Note that most of our viable models have DM masses above a few MeV,
which is in excellent agreement with bounds on big-bang nucleosynthesis
(BBN) \cite{Depta:2019lbe}.

\subsection{Experimental sensitivity}

\begin{figure}[t] 
 \centering
 \includegraphics[width=0.8\textwidth]{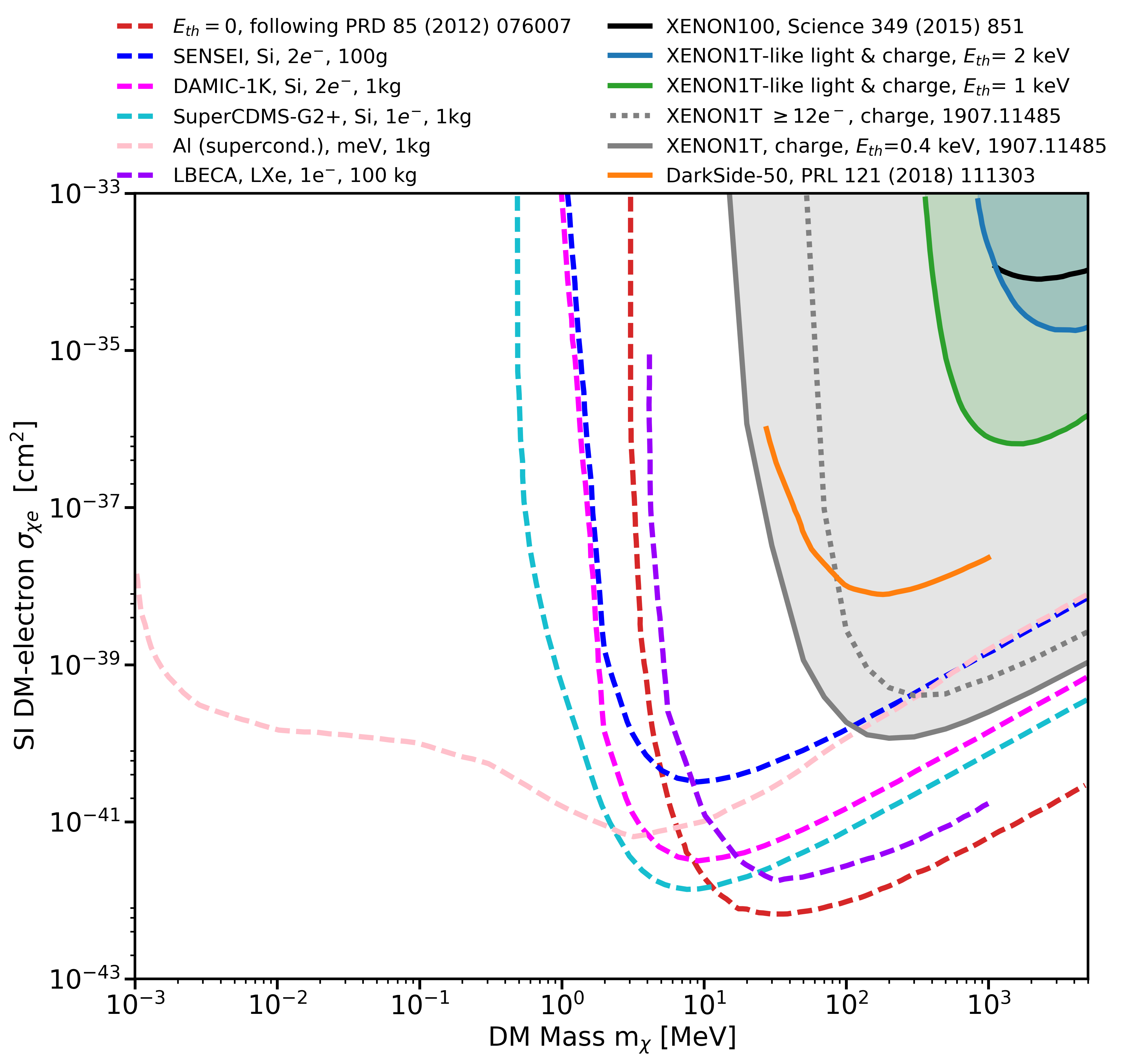}
 \caption{Experimental upper limits and estimated sensitivities for the
   DM-electron scattering cross section $\bar{\sigma}_{\chi \text{e}}$ in
   LXe, LAr and other detectors. Shown are the previously estimated
   sensitivity for a 60 kg-year exposure assuming only irreducible
   neutrino backgrounds and no detection thresholds (red dotted)
   \cite{Essig:2011nj} together with our sensitivity estimates for an
   assumed electron energy detection threshold of 2 keV (blue) and 1
   keV (green), respectively, for a 1 ton-year exposure of XENON1T
   with realistic assumptions at 95~\% confidence level (C.L.)
   \cite{Aprile:2018dbl}. Also shown are published limits from
   XENON100 (black) \cite{Aprile:2015ade} and a charge-only analysis
   of XENON1T \cite{Aprile:2019xxb} (gray full line), both at 90~$\%$
   C.L. The gray dotted line is a more conservative version of the
   latter limit considering that electron recoils below 186~eV (12 produced
   electrons) are undetectable, as the LXe charge yield $Q_y$ has never
   been measured below these energies \cite{Aprile:2019xxb}.
   For comparison, we also show the ionisation-only limit in Argon from
   DarkSide-50 (orange full line) \cite{Agnes:2018oej} and sensitivity
   projections (dashed lines) for SENSEI (blue), DAMIC-1K (magenta), Al
   (supercond.) (pink), SuperCDMS (cyan) \cite{Battaglieri:2017aum} and
   LBECA (purple) \cite{lbeca}.
 }
\label{fig:7}
\end{figure}

For DM with sub-GeV mass and no nuclear interactions, direct detection
experiments must rely on the recoil of (one or several) electrons
leading to ionisation. The cross section sensitivity of an experiment
with a liquid xenon (LXe) target and an exposure of 1 kg-year, assuming
a constant form factor, no detector threshold and only the irreducible
neutrino background, was initially estimated to be a few times $10^{-41}$
cm$^2$ for masses of a few tens of MeV \cite{Essig:2011nj}. In Fig.
\ref{fig:7} we plot this sensitivity (red dotted line) for an exposure of 60
kg-year by simply scaling it down by a factor of 60, neglecting the
influence of the tiny but non-zero irreducible background of neutrino
scattering. We will see below that an exposure of 60
kg-year can be realistically achieved with present experiments. This
sensitivity is, however, still ten orders of magnitude away from
the cross section range shown in Fig. \ref{fig:4}. 

In the following, we investigate how realistic the assumptions of no
detector energy threshold and only irreducible neutrino background for
DM experiments with a LXe target are, taking the example of XENON1T with
the so far largest LXe target. Its ultra-low background rate for
electron recoils of $82^{+5}_{-3}$ events/(ton$\cdot$year$\cdot$keV) at
low energies represents the lowest background level in the world for a
DM experiment. It is achieved by a strict material selection w.r.t.\
radio-purity and by fiducialising an inner self-shielded LXe volume. The
interaction point is reconstructed by using both the scintillation light
(S1) and the charge signal (S2). Unfortunately, demanding an S1 signal 
restricts the electron recoil energy to more than 2.5 (2.3) keV
for XENON10 (XENON100) \cite{Baudis:2013cca} and 1.3 keV for XENON1T
\cite{Aprile:2018dbl}. These energies are significantly larger than the
electron binding energy of a few tens of eV.
First, we estimate the sensitivity of the XENON1T experiment for the 1
ton-year exposure collected for the spin-independent WIMP nuclear
recoil analysis \cite{Aprile:2018dbl}, using realistic backgrounds and
energy thresholds and the standard mode of fiducialisation with both the
S1 and S2 signals. As discussed in Ref.\ \cite{Aprile:2015uzo}, the
electron recoil background in XENON1T is expected to be flat in the
region of interest (ROI) of the WIMP search and dominated by
$\beta$-decays of the $ ^{214}$Pb originating from $ ^{222}$Rn
emanations. The largest sub-dominant background comes from $\beta$-decays of
$ ^{85}$Kr, whose average natural concentration is reduced through cryogenic
distillation down to the sub-ppt level \cite{Aprile:2016xhi}. The energy
threshold is defined by a corrected scintillation (cS1) signal between 3
and 70 photoelectrons (PE) and an ionisation (S2) signal threshold of 200
PE \cite{Aprile:2018dbl}.
Similarly to the procedure in Ref.\ \cite{Aprile:2014eoa}, a conversion function
between S1 in PE and energy recoil in keV can be derived by using the Noble
Element Simulation Technique (NEST) \cite{Szydagis:2011tk}. As further
discussed in Ref.\ \cite{Aprile:2019dme}, NEST contains a global analysis of
LXe measurements from all available historical data for the mean
photon and charge yield, including XENON1T. Thus, a reliable conversion down
to the 1 keV energy threshold, where the detection efficiency of XENON1T drops
to less than 10$\%$, is available. Using the NEST conversion function, we
explore both a conservative detection threshold of 2 keV, where a full
triggered efficiency is still possible, and for illustration also 1 keV,
where the detection efficiency is already limited. Using Poisson statistics
(cf.\ Tab.\ XXII of Ref.\ \cite{Feldman:1997qc}), we then obtain for each
assumed DM mass an average upper limit on the event rate and thus the
experimental sensitivity on the reference cross section
$\bar{\sigma}_{\chi \text{e}}$ (green and blue lines in Fig. \ref{fig:7}).
In principle, the sensitivity can be improved by using the expected annual
modulation of a DM signal in an earth-bound detector rotating once per year
around the sun and thus possessing an annually modulated velocity w.r.t.\
to the DM wind. Using this time information, the XENON100 experiment with
its electron energy threshold of 2.3 keV became sensitive to DM masses down
to 600 MeV with a lowest cross section of $6\cdot10^{-35}$ cm$^2$ for a DM
mass of 2 GeV and axial-vector couplings \cite{Aprile:2015ade} (black line
in Fig. \ref{fig:7}).

To really come closer to the ideal sensitivity line, the energy threshold
has to be lowered drastically by abandoning the scintillation light
requirement S1 and using only the charge signal S2, which has a lower energy
threshold. The disadvantage of this method is that fiducialisation is limited,
since without S1 the event depth $z$ cannot be accurately estimated, yielding
a potentially increased background. Furthermore, there are additional
single-electron backgrounds coming from the photoionisation of impurities of
the LXe and the metal components inside the time-projection chamber
\cite{Aprile:2013blg}, which are difficult to describe without considering the
information provided by the scintillation light. Therefore, a reliable
background model becomes difficult. With this method, searches for DM scattering
off nuclei with XENON10 with a 1 keV nuclear recoil energy threshold and an
exposure of 15 kg-days \cite{Angle:2011th} as well as with XENON100 with a
0.7 keV nuclear recoil energy threshold and an exposure of 30 kg-years
\cite{Aprile:2016wwo} have been performed, assuming very conservatively that
all counts in the search window are potential signal counts. In Ref.\
\cite{Essig:2011nj} these analyses were re-interpreted in terms of light DM
scattering off electrons, and exclusion limits of $3\cdot10^{-38}$ cm$^2$ or
just below $10^{-38}$ cm$^2$ at 100 MeV were obtained.

Very recently XENON1T published a S2-only analysis \cite{Aprile:2019xxb}
with an electron energy threshold of 0.4 keV. Strict cuts on the data
were applied to lower the background drastically down to 1 event/(ton
$\cdot$day$\cdot$keV), resulting in a remaining effective exposure of
22 ton-days or 60 kg-years. The analysis considered a flat electron
recoil background originating mainly from $ ^{214}$Pb decays, coherent
elastic neutrino-nucleon scattering and ``cathode events'', which were
attributed to low energy $\beta$-electrons from the cathode wire grid.
This analysis was sensitive to light DM scattering off electrons down
to masses of 100 to 20 MeV (gray lines in Fig. \ref{fig:7}).

Comparing the estimated sensitivities and experimental limits plotted in
Fig. \ref{fig:7} to the reference cross section of at most a few times
10$^{-52}$ cm$^2$ in the SLIM model in Fig. \ref{fig:4}, we observe a
difference of  ten orders of magnitude despite the MeV to GeV DM mass
and large electron couplings. The reason is of course the large
suppression by the fourth power in the mass of the charged heavy mediator.
Despite the fact that XENON1T has demonstrated unprecedented low background
levels, the exclusion reach is still heavily constrained by the detection
threshold. Skipping the requirement to detect the scintillation light
signal S1 together with the charge signal S2 allowed to lower the energy
threshold already by one order of magnitude down to 0.4 keV and the
cross section limit by many orders of magnitude. Cryogenic bolometers can
have even lower energy thresholds (e.g. CRESST-III \cite{Abdelhameed:2019hmk}),
but they suffer from much smaller detector masses.
In addition, alternative technologies such as the charge-coupled-device (CCD)
experiments SENSEI \cite{Abramoff:2019dfb} and DAMIC \cite{Aguilar-Arevalo:2016zop}
with 0.1 kg to 1 kg (1K) target mass, respectively, a low-threshold
Generation-2 (G2+) SuperCDMS detector \cite{Agnese:2015nto} or a
superconducting Aluminum cube \cite{Hochberg:2015fth} are currently being explored
\cite{Battaglieri:2017aum}. The liquid xenon experiment LBECA also aims
at a significant background reduction for single or few electron signals \cite{lbeca}.
Note that the current limits, e.g.\ from the
first shallow underground run of SENSEI, are still eight orders of magnitude
weaker than the projected sensitivity  \cite{Abramoff:2019dfb}.
The projections are therefore shown in Fig.\ \ref{fig:7} as dashed lines.


\section{Conclusion}
\label{sec:7}

Motivated by the observations of DM, fundamental scalars and non-zero
neutrino masses in Nature, we have studied in this paper right-handed
neutrinos with MeV to GeV mass, that can explain the DM and generate
the active neutrino masses at one loop through their couplings to the
SM neutrinos via a dark scalar doublet, mixing with a scalar singlet.
When the mass of the lightest neutral scalar is also as light as MeV
({\it i.e.} a so-called SLIM), this scotogenic model, as had been observed
previously, can in addition solve the cosmological missing satellite,
cusp-core and the too-big-to fail problems.

We first scanned the full parameter space with right-handed neutrino
DM in the MeV to GeV mass range to identify viable models that reproduce
the observed SM neutrino mass differences and lead to the correct observed
relic density. We found that these models automatically have DM masses
above a few MeV in excellent agreement with bounds from big-bang
nucleosynthesis. We also observed that they implied sizeable couplings to
not only the neutral, but also the charged SM leptons between a few times
10$^{-4}$ and a few times $10^{-1}$.

We then turned to lepton flavour violation, which had previously not
been analysed, but which occurs naturally in this type of models due to the
coupling of the dark sector to both the charged and neutral SM leptons.
Specifically, we calculated the expected branching ratio of the process
$\mu\to e\gamma$, which is usually the most sensitive one. This process
turned out to be highly constraining, with the current MEG results already
eliminating a substantial part of the lepton coupling parameter space.
Since the latter is constrained from below from the requirement of non-zero
active neutrino masses, the planned MEG upgrade should almost completely
verify or exclude this model. The current limits on the processes $\mu\to3e$
and muon to electron conversion in Titanium proved to be less constraining,
but the next generation of experiments should be of similar sensitivity.

Since no nuclear recoil signal was expected for this leptophilic DM, we
then investigated instead the possibility of an electron signal,
which had generally been suggested as a possible way to detect sub-GeV DM.
We found that, despite the sizeable coupling of the right-handed neutrino
DM to charged leptons, the electron recoil cross section did not exceed a few
times $10^{-46}$ cm$^2$, when all low-energy, collider, cosmological and
neutrino mass and mixing constraints were imposed. This result could be
traced to the fact that the charged scalar mediator was restricted by LEP
searches to be heavier than 98.5 GeV and suppressed the cross section with
the fourth power of the mass. Lepton flavour violation constraints limited
the electron recoil cross section further down to at most $10^{-52}$ cm$^2$.
We furthermore performed a detailed study of the sensitivity of XENON and
other direct detection experiments, in particular to DM induced
electron recoil for a 1 ton-year exposure collected with realistic
backgrounds and energy thresholds, and compared it to previous theoretical
estimates without backgrounds or threshold limitations. Here, we found that
cross sections of about $10^{-36}$ cm$^2$ rather than a few times $10^{-41}$
cm$^2$ and DM masses of a few 100 MeV rather than a few tens of MeV were
required. Furthermore, we included the recent published S2-only analysis
result from XENON1T that features a lower electron energy threshold of 0.4
keV and a simplified, but realistic background model. The cross sections of
about $10^{-39}$ cm$^2$ for DM masses a few tens of MeV demonstrated several
orders of magnitude of improvement in the XENON sensitivity, approaching
the optimistic theoretical estimates. This makes it imperative to increase
the exposure in the future at a threshold below the keV level,
{\it e.g.} by exploiting further only the ionisation (S2) signal. There,
XENONnT will rival the sensitivities of upcoming or planned dedicated
experiments exploring the MeV mass range such as SENSEI, DAMIC-1K, SuperCDMS-G2+
or superconducting Aluminum cubes. We cautioned again that theoretical
projections may be overly optimistic, as they were in the case of XENON,
since e.g.\ the proven sensitivity from the first underground run of SENSEI
is still eight orders of magnitude below the projection.

We close with two remarks on possible generalisations of this model and
future directions of research. Indeed, some of our findings do not depend
on the details of the model. In particular, our results should apply more
generally to MeV to GeV scale neutrino DM and depend neither on the details
of the symmetry stabilising the DM nor on the presence of an equally light
scalar, as the main feature was the coupling of MeV neutrino DM to SM
leptons via the scalar doublet and the LEP limit on the mass of its charged
component. In contrast, promoting the
$U(1)$ symmetry stabilising the DM to a local symmetry and then studying the
phenomenology of its spontaneous breaking at colliders and in cosmology
would represent an interesting future direction of research
\cite{Klasen:2016qux}.


\section*{Acknowledgements}

We thank J.\ Alvey for useful discussions.
This work has been supported by the BMBF under contract 05H18PMCC1 and the
DFG through the Research Training Group 2149 ``Strong and weak interactions
-- from hadrons to dark matter''.


\end{document}